\documentclass[aps,prl,preprint,groupedaddress,showpacs,amssymb,amsmath]{revtex4-1}

\usepackage{graphicx}

\begin{document}


\title{Modified Coulomb potential and the S-state wavefunction of heavy quarkonia}


\author{Shwe Sin Oo}
\email[Email:]{shwe.oo@usm.edu}
\author{Sungwook Lee}
\email[E-mail:]{sunglee@usm.edu}
\author{Khin Maung Maung}
\email[Email:]{khin.maung@usm.edu}
\affiliation{School of Mathematics and Natural Sciences, The University of Southern Mississippi, Hattiesburg,
MS 39401, USA}



\begin{abstract}
The combination of relativistic kinematics together with Coulomb or Yukawa potentials is a common place in atomic, nuclear and meson mass phenomenology. In meson spectroscopy, linear and Coulomb-like potentials are the most commonly used potentials and decays rates are also calculated using the value of the wavefunction
at the origin. But, it has been known for sometime that the use of relativistic kinematics together with Coulomb or Coulomb-like potentials produces wavefunction singularity at the origin for S-wave \cite{DuranFriarPolyzou}.   
Therefore, there is a need to address this problem somehow.  In this letter, we show how to overcome this problem without destroying the phenomenological success of the use of relativistic kinematics in meson mass spectroscopy.
Our method can be easily implemented for similar problems in atomic, nuclear and particle phenomenology.
\end{abstract}


\maketitle

\section{Introduction}
Relativistic kinematics and Coulomb or Coulomb-like potentials are frequently used in atomic physics, nuclear physics and particle phenomenology. But, it was originally pointed out by Duran and Duran and later confirmed by Friar and Polyzou \cite{DuranFriarPolyzou} that, with relativistic kinematics and Coulomb potential, the S-wave Schr\"{o}dinger wavefunction is singular at the origin. This is true for all potentials with $1/r$ fall off such as the Coulomb and Yukawa potential.

In meson spectroscopy, two main potentials that are in use are the linear potential and the Coulomb-like potential.  The linear potential is to ensure the quark confinement and the Coulomb-like potential is to simulate asymptotic freedom. These potentials have been used with or without the relativistic kinematics and meson mass spectra in most part are well described. Once the fit to the meson mass spectra is done, one can calculate decay rates of these mesons from the associated wavefunctions. In simplest terms, decay rate is proportional to the square of the wavefunction at the origin. Therefore, in doing meson mass spectroscopy, it is important to get the correct wavefunction. There is no obstacle as long as non-relativistic Schr\"{o}dinger equation is concerned. The problem appears when one wants to employ relativistic kinematics. 

As far as we are aware, none of the relativistic calculations take this wavefunction singularity at the origin into account. Because of the radical sign associated with relativistic kinematics, all calculations resort to an expansion of the wavefunction in some basis and reduce the Schr\"{o}dinger equation into a matrix eigenvalue equation. We note that, since one cannot use an infinite number of basis functions in numerical calculations, in principle, these calculations are variational calculations. Because of the nature of variational calculation, the mass spectrum can be described quite well, although the wavefunction might not reflect all the properties of the Hamiltonian \cite{Messiah1958}.  A basis function set which includes this singularity exists, but it was pointed out that a naive application can produce unphysical predictions \cite{Polyzou1997}.

In our opinion, this wavefunction sigularity is a pathology arising from the combination of relativistic kinematics and Coulomb-like potential and not physical. Jarome Malenfant \cite{Malenfant87} showed that in QED, this divergence goes away when higher order diagrams are considered and conjectured that this will also be the case in QCD. H. Ito \cite{HIto1987} studied a related problem for the Dirac equation and showed that after renormalization, the wavefunction singularity can be made to disappear. In this paper, we will directly attack the problem by modifying the potential. Since the behavior of the Coulomb-like potential near the origin will mostly affect on the low lying states of the heaviest of quarkonia states, we will introduce a modification of the Coulomb-like potential in such a way that it will not affect the phenomenology.

We will demonstrate that using the usual linear plus the unmodified Coulomb-like potential together with relativistic kinematics for bottomonium, the $1 S$ state mass converges as the number of basis functions is increased, but the value of the wavefunction at the origin does not. This shows that decay rates using these wavefunctions are unreliable. We will also demonstrate that a simple modification of the Coulomb-like potential can fix this problem.
\section{Modification of the potential and method of calculation}
Because of the radical sign, usually, relativistic calculations are performed in momentum representation using momentum space basis functions. Although, both the linear potential and the Coulomb potential are singular in momentum representation, methods to deal with these singular potentials are well known \cite{VaryMaung}. Obviously, the singularity of the wavefunction at the origin stems from the fact that Coulomb potential is singular at the origin. What we propose to do is to modify the Coulomb potential in such a way that the wavefunction singularity problem is solved and the usual success of phenomenology is maintained. To this end, we propose our modified Coulomb-like potential as
\begin{equation}
V_{Coul}(r)=- \frac{C}{r+\epsilon}
\end{equation}
where $C$ and $\epsilon$ are positive constants. Obviously, in order to keep the phenomenological success, $\epsilon$ has to be small. 

We consider the following Hamiltonian 
\begin{equation}
H=A({\bf p}) +\sigma r - \frac{C}{r+\epsilon}+C e^{-\frac{r^2}{4r_0^2} }\frac{{\bf \sigma_1 \cdot \sigma_2}}{3m^2r_0^3\sqrt{\pi}}
\end{equation}
where
$$A({\bf p})=\sqrt{ {\bf p}^2 + m_1^2} + \sqrt{ {\bf p}^2 + m_2^2}$$
Here $m_1$ and $m_2$ are the quark and antiquark  masses respectively, $\sigma$ is the strength of the linear potential, and the last term is the spin-spin potential. Since we are interested only in S-wave, we do not need to consider the spin-orbit interaction. In order to deal with the radical sign and the modified Coulomb potential together at the same time, we make use of the expansion of the  wavefunction in both position space and also in momentum space. We can expand the momentum space wavefunction and the position space wavefunction \cite{Polyzou1997} as 
\begin{align}
{\tilde \psi}({\bf p})&=\sum\limits_{nlm}{\tilde C}_{nlm}~{\tilde \phi}_{n}^{l}(p)~Y_l^{m}({\hat p})\\
{\psi}({\bf r})&=\sum\limits_{nlm}C_{nlm}~{ \phi}_{n}^{l}(r)~Y_l^{m}({\hat r})
\end{align}
The position space basis functions we use \cite{Polyzou1997} are linear combinations of products of Laguerre polynomials and an exponential function, and the momentum space basis functions are Jacobi polynomials. They are related to each other by a Fourier-Bessel transformation
\begin{align}
\phi_{n}^l (r)&=\sqrt{ \frac{2}{\pi}} \int\limits_0^{\infty} j_l(pr) {\tilde \phi}_n^l(p) p^2 dp\\
{\tilde \phi}_{n}^l (p)&=(i)^l~\sqrt{ \frac{2}{\pi}} \int\limits_0^{\infty} j_l(pr) {\phi}_n^l(r) r^2 dr
\end{align}
We can then convert the Schr\"{o}dinger equation into a matrix eigenvalue problem. The only requirement is that the position space basis functions and momentum space basis functions are related by a Bessel transformation\cite{Polyzou1997} \cite{SSONewOrleans}.  The eigenvalue equation is given by
\begin{equation}
\label{eq:eigenvalue}
\sum\limits_n C_{nlm} \bigg[ \int\limits_0^{\infty} {\tilde \phi }_{n'}^l (p) A(p) {\tilde \phi}_n^l(p) p^2dp +\int\limits_0^{\infty}\phi_{n'}^l(r)V(r)\phi_n^l(r)r^2dr \bigg]=E_{nl}C_{n'lm}
\end{equation}
We note that the equation \eqref{eq:eigenvalue} handles the potential in position basis and the kinematics part in momentum basis. Hence, it can deal with any form of potential including non-integer power law potentials and at the same time we can use relativistic kinematics. 
\section{Calculations and Results}
Since we are modifying the Coulomb-like potential for small distances, we expect that it will affect the lowest lying states of heavy mesons such as $b{\bar b}$ and $c{\bar c}$. To be concrete, we will study the $b{\bar b}$ system for $l=0$ case. Since we are dealing with S-states, we will not need to consider the spin-orbit potential. This will be included for $l\neq 0$ cases in a later study. 

\begin{figure}[h]
	\includegraphics[height=4.0in]{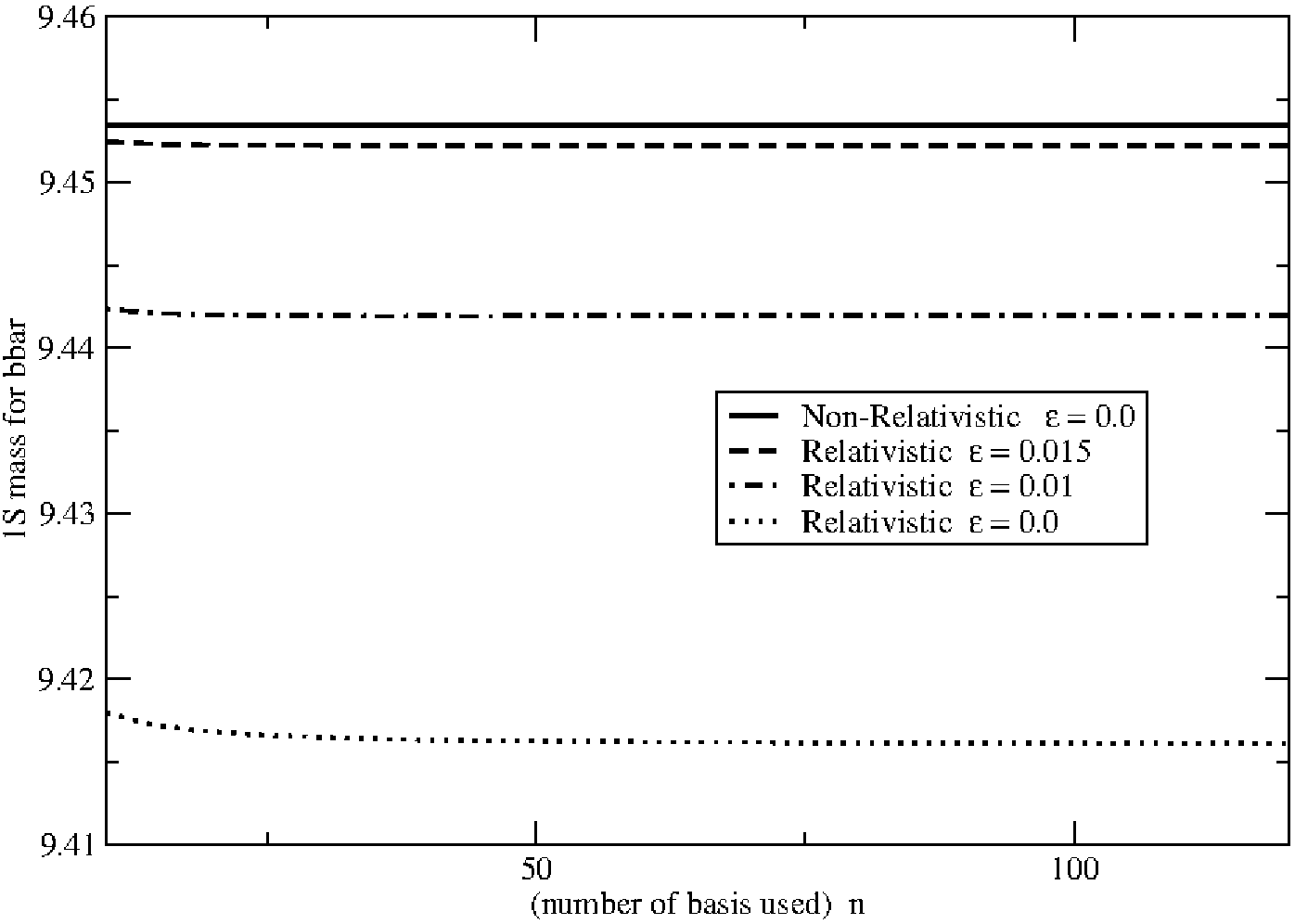}
	\caption{Mass of 1S state vs number of basis functions used. $m_b=4.77~GeV$, $\sigma=.197 ~GeV^{2}$, $C=0.5$.\label{fig:figure1}}
\end{figure}

First, we show the results with only linear plus Coulomb-like potential (without and with modification). We use the linear potential strength $\sigma=0.197~ GeV^{2}$ and Coulomb strength $C=0.5$ and quark mass $m_b=4.77 ~GeV$. In Fig. \ref{fig:figure1}, the mass of the  $1S$ state is plotted against the number of basis functions used in the expansion of the wavefunction. The non-relativistic curve converges rapidly as the number of basis increases. The  relativistic curve with unmodified Coulomb ($\epsilon = 0.0$) has the slowest convergence, but all relativistic curves converge to their respective final values. 

\begin{figure}[h]
	\includegraphics[height=4.0in]{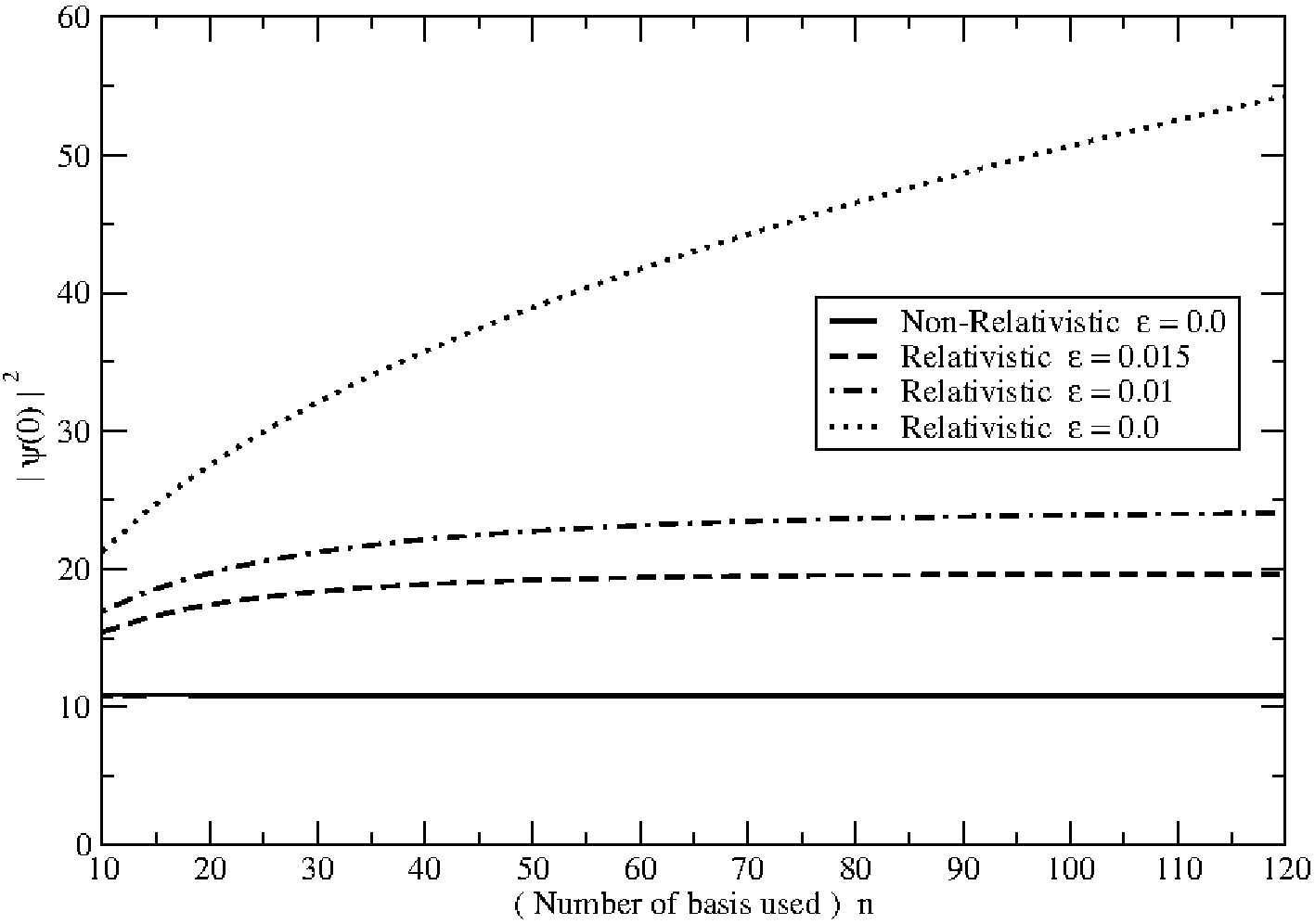}
	\caption{$| \psi(0)|^2$ vs number of basis functions used. $m_b=4.77~GeV$, $\sigma=.197 ~GeV^{2}$, $C=0.5$.\label{fig:figure2}}
\end{figure}

Next, in Fig. \ref{fig:figure2}, we show the value of the squared wavefunction at the origin $|\psi(0)|^2$ against the number of basis used. Again, the non-relativistic curve converges rapidly, but in the relativistic cases, we see that the unmodified ($\epsilon =0.0$) case does not converge at all. As we can see, $| \psi(0) |^2$, the squared value of the wavefunction at the origin diverges as the number of basis functions is increased.  This is an important result and it shows that decay rate calculations with unmodified Coulomb can lead to discrepancies depending on the number of basis functions used. We see that the results obtained with $\epsilon=0.015$ achieve convergence by the time the number of basis functions reaches $n=80$. Now, we have an unambiguous  way of determining $|\psi(0)|^2$ by introducing a parameter $\epsilon$ in the potential.

Finally, in Table \ref{tb:data}, we show that our model can still perform well phenomenologically also. The table shows the $S$ state $b\bar b$ masses compared to experimental values.
\section{Conclusion}
We have done calculations with relativistic kinematics using linear plus unmodified Coulomb potentials. We found that although the bound state mass seems to converge as we increase the number of basis functions, the squared value of the wavefunction at the origin $|\psi(0)|^2$ does not reach a stable value. Therefore, looking at the convergence of the eigenvalue (meson mass) alone is not sufficient, and one must also check the convergence of the wavefunction at the origin. Modifying the Coulomb potential gives convergence to both the masses and $|\psi(0)|^2$. In order to show that modification of the potential does not destroy the phenomenological success, we calculated the $b{\bar b}$ spectrum for $l=0$. The last column shows our calculated masses together with the percentage error in comparison with the experimental results. Our results agree well  with experimental results \cite{PDG2024}. The method we used can be easily implemented for similar problems in atomic, nuclear and particle phenomenology.
\begin{acknowledgments}
We would like to thank Dr. Parthapratim Biswas for useful discussions. 
\end{acknowledgments}

\bibliography{modified-coulomb.bib}
\newpage
\newpage
\newpage

\begin{table}
\begin{tabular}{|c|l|l|l|l|}
	\hline
	\hline
	~&~&~&Expt.&This work\\
	Name& $n ~^{2s+1}L_J$&$ I^G (J^{PC})$& Mass(GeV) &(percent error)\\
	~&~&~&PDG&~\\
	\hline
	\hline
	~&~&~&~&\\
	$	\eta_b$&$ 1 ~^1S_0$&$ 0^+ ( 0^{-+})$& $9.3910 $&9.3930\\ 
	~&~&~&~&(0.021\%)\\
	\hline
	~&~&~&~&\\
	$\eta'_b$& $2 ~^1S_0$&$ 0^+ ( 0^{-+})$& $9.999$&10.0001\\ 
	~&~&~&~&(0.011\%)\\
	\hline
	~&~&~&~&\\
	$\Upsilon$&$ 1 ~^3S_1$&$ 0^- ( 1^{--})$& $9.4602 $&9.4603\\ 
	~&~&~&~&(0.0011\%)\\
	\hline
	~&~&~&~&\\
	$\Upsilon^{'}$&$ 2 ~^3S_1$&$ 0^- ( 1^{--})$& $10.0232$&10.0257\\ 
	~&~&~&~&(0.025\%)\\
	\hline
	~&~&~&~&\\
	$\Upsilon^{''}$&$ 3 ~^3S_1$&$ 0^- ( 1^{--})$& $10.3552$&10.3778\\ 
	~&~&~&~&(0.218\%)\\
	\hline
	~&~&~&~&\\
	$\Upsilon^{'''}$&$ 4 ~^3S_1$&$ 0^- ( 1^{-1})$& $10.5794$&10.6635\\ 
	~&~&~&~&(0.795\%)\\
	\hline
\end{tabular}
\caption{$m_b=4.7662~GeV$, $\sigma=0.197~ GeV^2$, $C=0.5$, $r_0=0.22 ~GeV^{-1}$,
$\epsilon ~=~0.015~GeV^{-1}$\label{tb:data}}
\end{table}
\end{document}